\documentclass[11pt, a4paper, twocolumn]{article}
\usepackage{inputenc}
\usepackage{hyperref}
\usepackage{amsmath}
\usepackage{physics}
\usepackage{amsfonts}
\usepackage{multirow}
\usepackage{color, soul}
\usepackage{orcidlink}
\usepackage{graphicx}
\usepackage{abstract}
\usepackage[affil-it]{authblk}
\usepackage[sorting=none, style=numeric-comp]{biblatex}
\bibliography{main}

\hypersetup{pdftitle={Convolutional Neural Network based Decoders for Surface Codes}}

\title{Convolutional Neural Network based Decoders for Surface Codes}

\author[1]{Simone Bordoni \orcidlink{0000-0002-9745-9189}}
\author[1]{Stefano Giagu \orcidlink{0000-0001-9192-3537}}
\affil[1]{La Sapienza University of Rome, Dep. of Physics, Rome, Italy.}

\begin{document}
\twocolumn[
\maketitle
\begin{@twocolumnfalse}
\begin{abstract}
The decoding of error syndromes of surface codes with classical algorithms may slow down quantum computation. To overcome this problem it is possible to implement decoding algorithms based on artificial neural networks.
This work reports a study of decoders based on convolutional neural networks, tested on different code distances and noise models. The results show that decoders based on convolutional neural networks have good performance and can adapt to different noise models. Moreover, explainable machine learning techniques have been applied to the neural network of the decoder to better understand the behaviour and errors of the algorithm, in order to produce a more robust and performing algorithm.\\ \vspace{10mm}
\end{abstract}
\end{@twocolumnfalse}]

\section{Introduction}

\begin{figure*}
    \centering
    \includegraphics[width=0.7\textwidth]{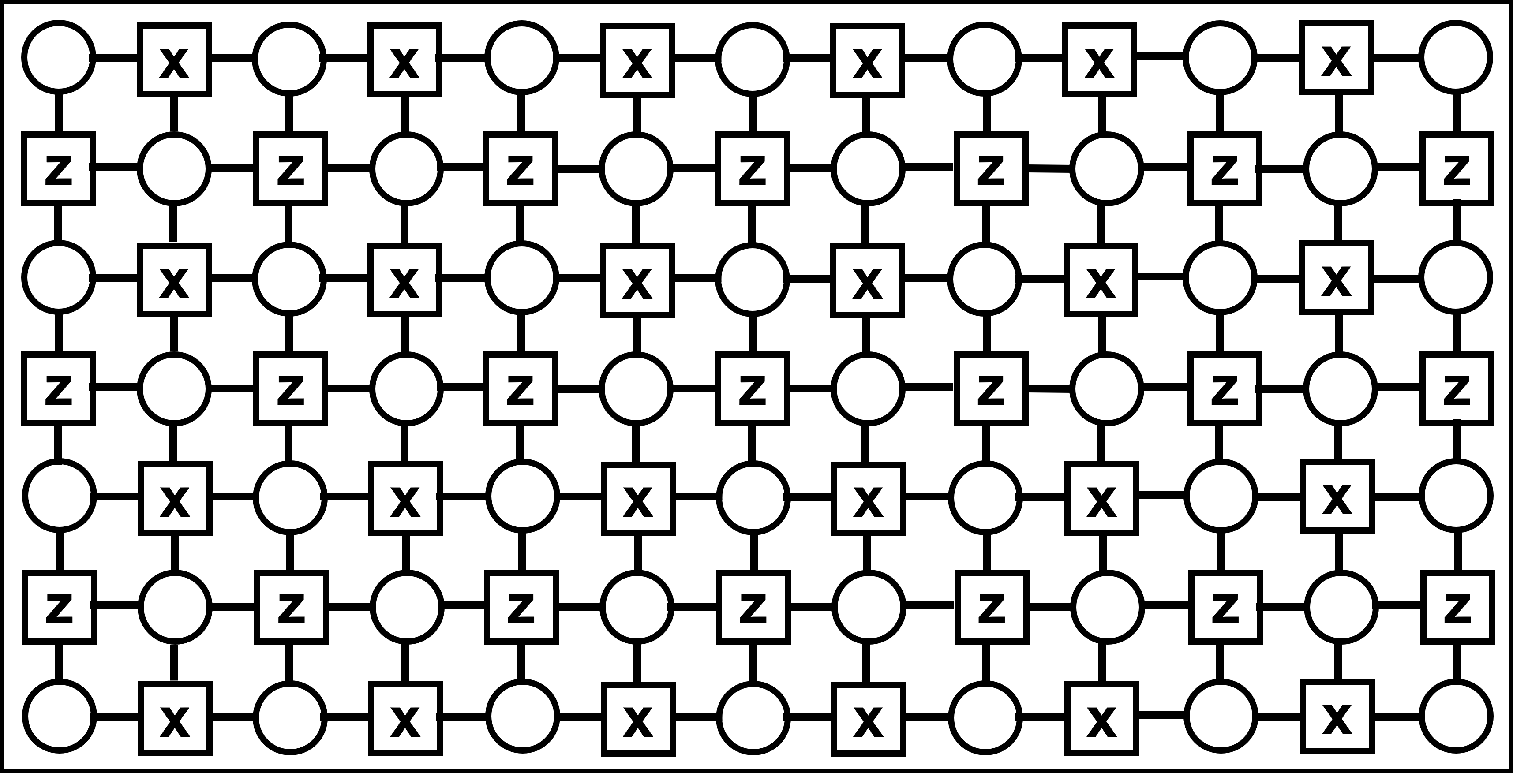}
    \caption{Schematic representation of a surface code. Data qubits are represented as circles while $X$ and $Z$ measurement qubits are represented as squares. Each qubit is connected with its nearest neighbours.}
    \label{fig:surface1}
\end{figure*}

In recent years a lot of interest has grown around the possibility of constructing efficient quantum computers. One of the main problems regards the protection of quantum information from external noise. This process, known as decoherence, is due to the inevitable interaction between the qubits and the environment~\cite{bib_decoherence, bib_principles2}. 
A solution may come from quantum error correction codes (QECC), where logical quantum information is stored in the degrees of freedom of a system composed of many physical qubits. In this way, the logical information can be restored even in the presence of single physical qubit errors~\cite{bib_qecc1, bib_qecc2}.
Surface codes are a family of QECC~\cite{bib_kitaev, bib_surface1}, their implementation is simpler, with respect to other codes, because only near interactions between the physical qubits is needed. In a surface code a greater number of physical qubits gives better protection from errors, however, this makes the decoding procedure more complicated. The decoding algorithm is implemented by a classical computer and should be carried out fast to not slow down the quantum computation. With classical algorithms it is difficult to obtain a reasonable execution time incrementing the number of physical qubits~\cite{bib_decodespeed}. A solution to this problem may come from the application of artificial neural networks.
Simple models of neural networks already give good decoding accuracy with a constant execution time for small surface codes~\cite{bib_vars0}. In order to decode codes with a larger number of qubits, more complicated neural network models have to be employed. This work focuses on decoders based on convolutional neural networks for different code distances, noise models and error probabilities.
To improve the performance of the algorithms different model architectures have been studied to find the best error probabilities for the training set and to reduce the number of trainable parameters of the neural network. In the last section of the article, to get a better understanding of the behaviour of neural network based decoders, explainable machine learning methods have been employed. An original technique to improve the decoder performance, based on data augmentation driven by the results of the model explainability, is also reported.

\section{Surface codes}\label{sec_surfacestab}

\begin{figure}[t]
\vspace{-0.15cm}
    \includegraphics[width=0.5\textwidth]{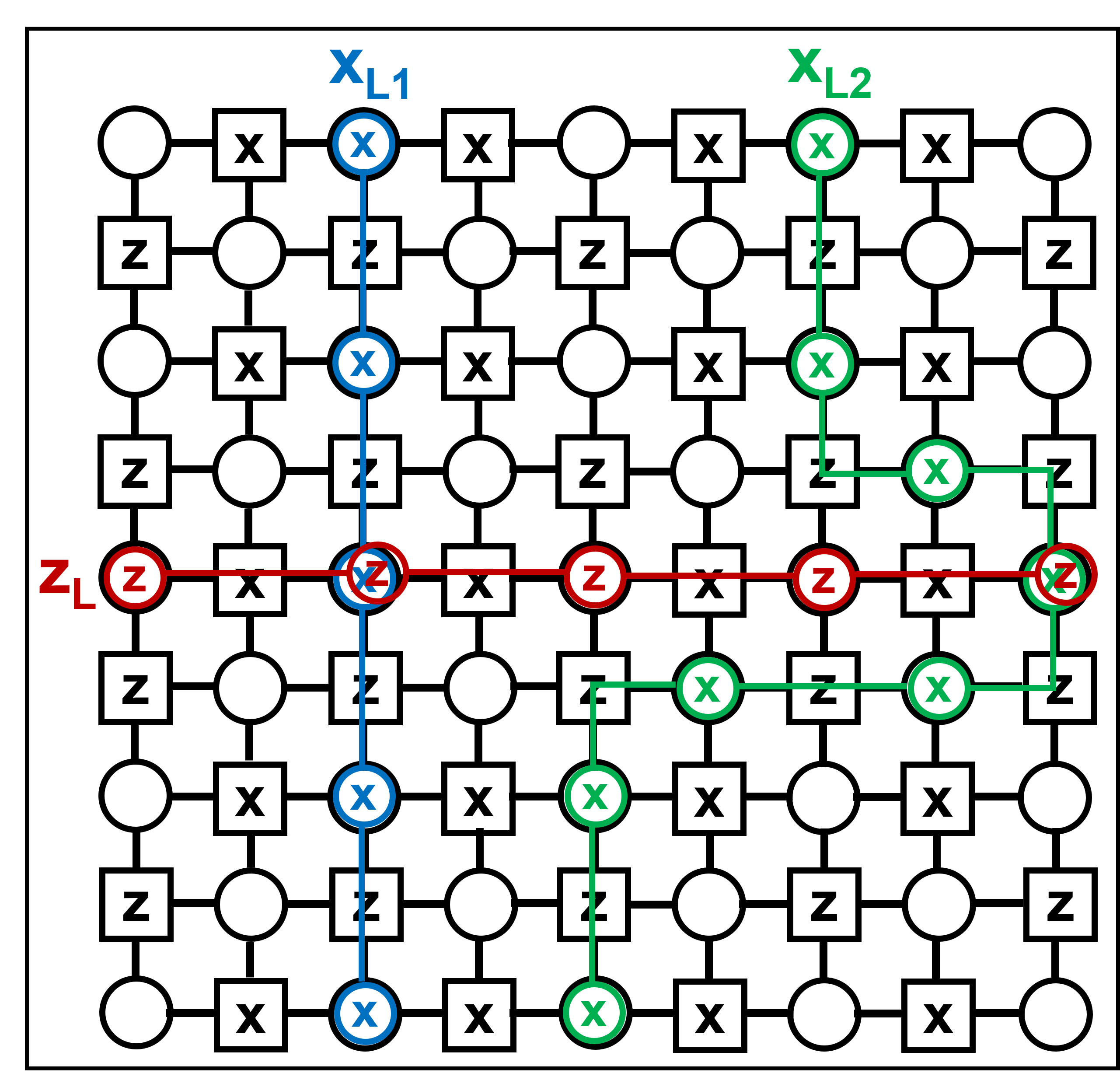}
    \caption{Examples of logical operators in a $d=5$ surface code. $Z$ and $X$ logical operators are composed of chains of single qubit $Z$ or $X$ operators connecting the $Z$ or $X$ sides respectively.}
    \label{fig:surface3}
\end{figure}

In a surface code physical qubits are arranged on a squared lattice; they are divided in two categories (Fig.~\ref{fig:surface1}). The \textit{data qubits} store the quantum information. The \textit{measurement qubits}, also called stabilizers, are used to perform projective measurements, on the nearest neighbour data qubits. Measurement qubits are divided into $Z$ and $X$ measurement qubits, they are used to perform respectively measurements of the Pauli operators $\sigma_x$ or $\sigma_z$~\cite{bib_qiqc}. To simplify the notation in the rest of this work $\sigma_x$ and  $\sigma_z$ Pauli operators will be called $X$ and $Z$ operators. As all the stabilizers commute, after a measurement cycle the system collapses into an eigenstate of all the stabilizers~\cite{bib_latticesurgery, bib_lattice}.
A surface code contains the logical information of a qubit. The logical state can be modified with logical operators. An operator of this kind has to preserve all the eigenvalues of the stabilizers~\cite{bib_logicalop}. On each border of a surface code there are only $X$ measurement qubits or $Z$ measurement qubits, these borders are called $X$ sides or $Z$ sides of the surface code respectively.
Every chain of single qubit $X$ operators, that connect the $X$ sides, works as an $X$ logical operator. This is also true, in the case of $Z$ logical operators, for chains of $Z$ single qubit operators that connect the $Z$ sides.
The minimum number of single qubit operators, necessary to change the logical state, is equal to the number of data qubits on a side of the lattice, this important value is called the distance of the code ($d$)~\cite{bib_surface1}. Figure~\ref{fig:surface3} shows some examples of logical operators in a $d=5$ surface code.

\subsection{Errors}\label{sec_noise}
\begin{figure}
    \includegraphics[width=0.5\textwidth]{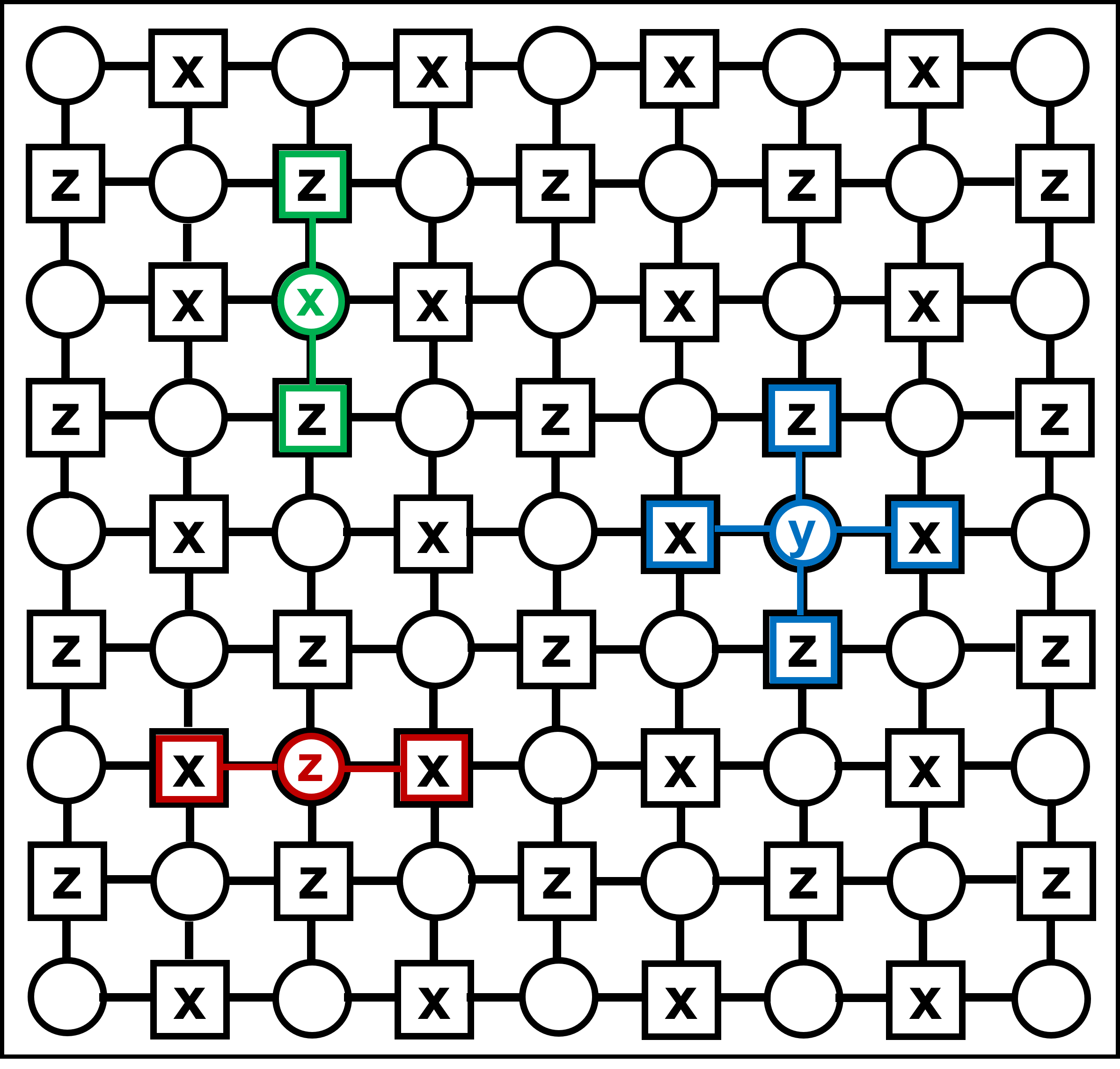}
    \caption{Signal error produced by single data qubit errors. The highlighted measurement qubits change their measurement value.}
    \label{fig:surface2}
\end{figure}

Unlike the classical bit, where only bit flip errors may appear, for a qubit there is a continuous set of possible errors. However, after a measurement cycle, the state of the qubits collapses into a discrete set of possible errors~\cite{bib_principles2, bib_qecc1, bib_qecc2}.
There are three main kind of errors that appear on a surface code.\\
\textit{Data qubit errors} are the classical bit flip ($X$ errors), phase flip ($Z$ errors) and a combination of both ($Y$ errors).
A single error of this type can be easily identified by observing a change of the eigenvalues in the neighbouring stabilizers (Fig~\ref{fig:surface2}).\\
\textit{Measurement errors} appear when the projective measurement fails. A single error of this type changes just the value of the considered stabilizer.\\
\textit{Gate errors} may appear during the measurement process when imperfect Hadamard and C-NOT gates are used. C-NOT gate errors are the most difficult to be identified because they affect two qubits.\\

It is possible to test different noise models (or error models) by including only some types of the errors previously described and by changing their probability. In this work many simulations of surface codes have been performed to test different code distances, error models and probabilities. For this purpose a specific python library has been employed\footnote{E. Villaseñor and B. Criger: https://github.com\\/evalvarez12/Distributed\_Surface\_Code.git}. The performance of a decoding algorithm can be very sensitive with respect to the considered error model and characteristics of the surface code~\cite{bib_noise1, bib_noise2, bib_noise3}.\\
In the \textit{depolarising error model} only data qubit errors are considered. Given $p$ the error probability, each data qubit is subject to an $X$, $Z$, or $Y$ error with equal probability $p/3$.\\ Perfect measurements and gates are assumed, thus only one measurement cycle is necessary.
In the \textit{circuit noise model} also the measurement qubits are subject to errors and gates are not assumed perfect. In this case it is better to make more measurement cycles to identify measurement errors.

\subsection{Decoding algorithms}
In a QECC, the Hilbert space that describes the state of the physical qubits can be decomposed into a logical and an ancillary subspace~\cite{bib_hilbert}.
After error correction the ancillary state, that increases the redundancy of information, is restored. This is not true for the logical state that may be modified, causing a logical error. The performance of a decoding algorithm can be tested by studying the rate at which logical errors appear, as a function of the single qubit error probability.\\
The minimum weight perfect matching algorithm (MWPM) restores the ancillary state with the minimum number of corrections~\cite{bib_mwpm1, bib_mwpm0, bib_mwpm2}. 
This decoder is only nearly optimal for bit-flip noise~\cite{XZopt} (independent $X$ and $Z$ errors), it represents a standard benchmark for other decoders.
When studying the logical error rate with respect to the single qubit error rate a typical behaviour occurs. For low error probabilities, the decoding accuracy increases with the dimension of the code. On the other hand, for high error probabilities, increasing the distance of the code reduces the accuracy of the decoder~\cite{bib_surface1, bib_noise2}. The cross-over of these two regimes occurs at the threshold error rate.\\

In order to use artificial neural networks, it is necessary to transform the original decoding problem in a classification problem. For this purpose, neural network based decoders are composed of two components:\\
The \textit{simple decoder} analyses the error syndrome and proposes a correction that matches with the syndrome. This algorithm may be implemented by a neural network but it is simpler to use naive decoder. In this work the simple decoder corrects each error with a chain of operators that connects it to the nearest border. In this way the ancillary state of the system is restored.\\
The \textit{high level decoder} (HLD) takes as input the error syndrome and tries to find out if the correction of the simple decoder has created a logical error. This is a classification problem that may be solved by a neural network.

\section{Related works}
Many different algorithms, based both on classical methods and machine learning techniques, have been tested for the decoding of surface codes. Apart from the MWPM previously introduced, examples of decoding algorithms not based on neural networks are the renormalization group decoder~\cite{bib_rg, rn2}, the cellular automaton~\cite{bib_ca}, the maximum likelihood decoder~\cite{bib_mld} and the Markov chain Monte Carlo decoder~\cite{bib_mcmc}. For these algorithms it is difficult to find a good compromise between decoding accuracy and execution time~\cite{bib_decodespeed}.
As neural network based decoders show a good compromise between accuracy and execution time, many studies have been carried out to test different neural network architectures. The first studies in this sector~\cite{bib_vars1} show that, for small surface codes, neural network based decoders have a decoding performance, similar to MWPM, for a depolarising noise model. When measurement error and imperfect gates are included it is possible to improve the MWPM algorithm taking into account of more complex noise models~\cite{optimal, corrxy}. However neural network based decoders remain interesting because they require constant execution time and can easily adapt to different noise models. Some problems arise for high distance surface codes as the number of possible error syndromes increases exponentially. This means that, for the correct training of the HLD, the training set have to increase with the dimension of the code to contain the most statistically relevant errors.\\

To scale the methodology to higher distance surface codes, some interesting approaches have been proposed. A recurrent neural network architecture has been tested by P. Baireuther et al. for the decoding of correlated errors~\cite{assisted}. G. Torlai and R.G Melko have tested a decoder based on a stochastic neural network (Boltzmann machine) that is applicable to a wide variety of stabilizer codes~\cite{boltzmann}. Other interesting studies on deep learning based decoders have been carried out by S. Krastanov and L. Jiang~\cite{toric}. S. Varsamopoulos et al.~\cite{bib_vars0} have compared decoders based on feed forward neural networks and on recurrent neural networks. Recent works have tested decoders based on distributed neural networks~\cite{bib_decodespeed}, in order to reduce the size of the training set for high distance surface codes. Another interesting novel idea comes from the application of machine learning techniques to an ensemble of classical decoders~\cite{ensemble}. Other decoders based on machine learning have been recently tested by D. Bhoumik et al.~\cite{bib_new2021} and C. Chamberland et al.~\cite{ch1, ch2}.
A state of the art decoder, scalable to high distance surface codes, has been created by K. Meinerz et al. combining convolutional neural networks (to preprocess local information) with a conventional algorithm~\cite{scalable}. Other decoders based on convolutional neural networks have been tested on high distance surface codes for a depolarising or bit flip error models~\cite{large1, large2}. While this work concentrates on smaller dimension surface codes, with respect to the previously mentioned studies, some new elements have been introduced. A more realistic noise model, that includes measurement errors, has been decoded with a convolutional neural network. The performance of the HLD have been studied with respect to the choice of the training set error probability. The dilated convolution technique have been tested to scale to higher distance codes and to reduce the number of trainable parameters of the neural network. Moreover, in order to understand better the cases where the neural network fails, explainable machine learning techniques have been applied to the HLD.

\section{Convolutional neural network based decoders}
In this section the characteristics of high level decoders based on feed forward neural networks (FFNN) and convolutional neural networks (CNN) will be tested. Two different noise models will be considered, the simple depolarising noise and a more complicated noise model where measurement errors and more measurement cycles are present. Before the comparison some preliminary studies are necessary for the tuning of the hyperparameters.
Furthermore, to improve the accuracy of the HLD, convolutional neural networks based on dilated convolution layers, able to better capture local features at different spatial scales, will be tested in this section.

\subsection{Depolarising noise}\label{sec_dep}
\begin{figure*}
\centering
\includegraphics[width=0.6\textwidth]{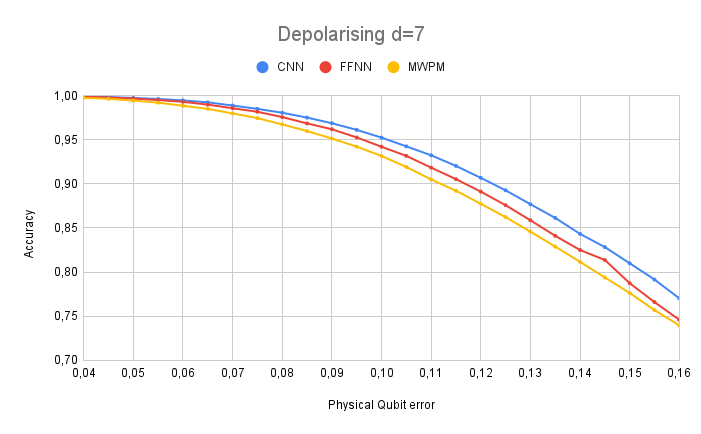}
\includegraphics[width=0.6\textwidth]{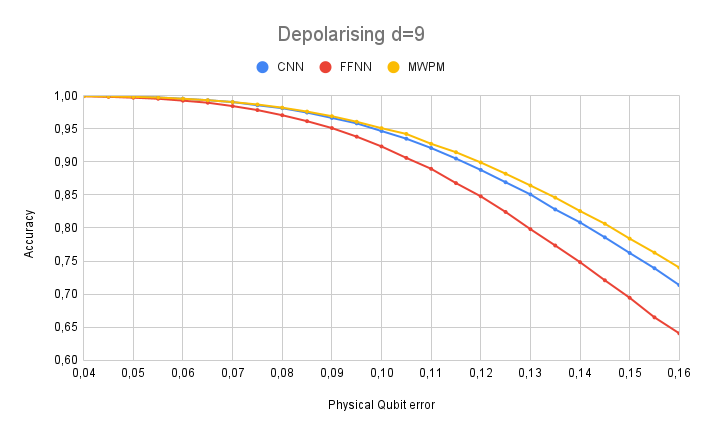}
\includegraphics[width=0.6\textwidth]{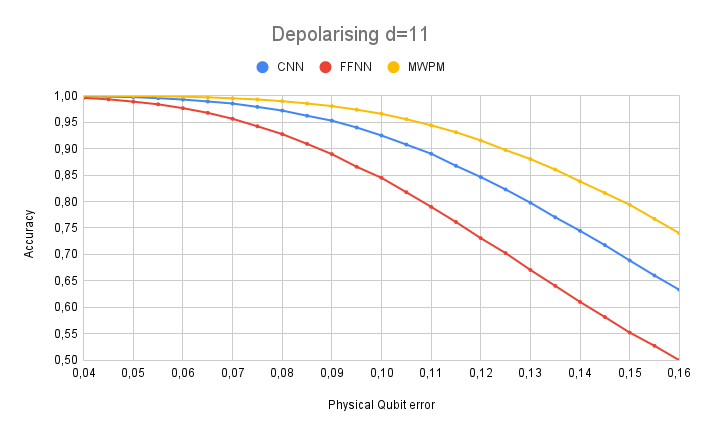}
\caption{Comparison between accuracy obtained with HLD based on FFNN and CNN and the accuracy obtained with MWPM for different code distances ($d$) and single qubit error probabilities (Physical Qubit error).}
\label{tab:arch2}
\end{figure*}

For the depolarising error model datasets of $5\times10^6$ elements have been generated with a single qubit error probability $p=0.1$ and for code distances $d=7,9,11$.
All the neural networks employed, both FFNN and CNN, share the following characteristics. ReLU activation functions have been used in all hidden layers while Softmax activation functions have been used for the output layer. The loss function employed is categorical cross entropy. The ADAM optimiser has been used, with a mini-batch size of 32 elements. The neural networks have been trained for 20 epochs using $10\%$ of the original dataset as evaluation set.
Feed forward neural networks with different numbers of layers and neurons have been tested on each dataset. The best results, for all the code distances, have been obtained using three hidden layers with a number of trainable weights reported in table~\ref{tab:p1}.
\begin{table}[h]
\centering
\caption{This table reports the optimal number of trainable weights ($P$) of the feed forward neural network obtained for different code distances ($d$) for a depolarising noise.}
\vspace{5mm}
\centering
\begin{tabular}{|c|c|}
\hline
$d$ & P \\
\hline
7 & $5.67\times 10^5$ \\
9 & $6.4\times 10^6$ \\
11 & $3.9\times 10^6$ \\
\hline
\end{tabular}
\label{tab:p1}
\end{table}\\
\noindent
For the convolutional neural networks, the input format consists of squared matrices with a number of rows and columns equal to $2 d-1$, where $d$ is the distance of the code. Each matrix element represents a qubit of the surface code. The value of the data qubits is set to zero, while the value of the measurement qubits is $\{1,-1\}$.
The first parameters of the CNN to be defined are the number of filters and the number of convolutional and dense layers. For the number of filters the best results have been obtained using $64$ kernels of dimension $3\times 3$.
The number of convolutional and dense layers that showed the best performance are reported in table~\ref{tab:arch}.
Other fine tests have been carried out to improve the performance, for example a stride with value $2$ has been tested for the first convolutional layer but this reduced the overall performance. A small increase in the accuracy has been obtained with a padding on the first layer.
The accuracy obtained with the previously described models of neural networks is reported in Fig.~\ref{tab:arch2} and has been compared to the accuracy obtained with MWPM for different code distances and single qubit error probabilities. The test on different error probabilities have been made on datasets of $2\times10^5$ elements.
For $d=7$ codes the performance of the high level decoder, both based on FFNN or CNN, is greater than MWPM. For greater distance codes the HLD has a worse performance than MWPM.
This is due to the fact that the dimension of the training set is still too small to include the most statistically relevant error syndromes, necessary for a correct training of the network. However, for these code distances, it is possible to observe the advantage in the use of a CNN with respect to a FFNN.

\begin{table}
\centering
\caption{This table reports best hyperparameters of the CNN for different code distances and a depolarising error model. $d$ is the code distance, CL is the number of convolutional layers, DL is the number of dense layers, N is the number of neurons in the dense layers and P is the total number of trainable weights.}
\vspace{5mm}
\centering
\begin{tabular}{|c|c|c|c|c|}
\hline
$d$ & CL & DL & N & P \\
\hline
7 & 3 & 1 & 512 & $2.7\times 10^6$ \\
9 & 4 & 1 & 1024 & $8.0\times 10^6$ \\
11 & 6 & 2 & 1024 & $9.2\times 10^6$ \\
\hline
\end{tabular}
\label{tab:arch}
\end{table}

\subsection{Depolarising plus measurement errors} \label{sec_ddep}

\begin{figure}
\includegraphics[width=0.49\textwidth]{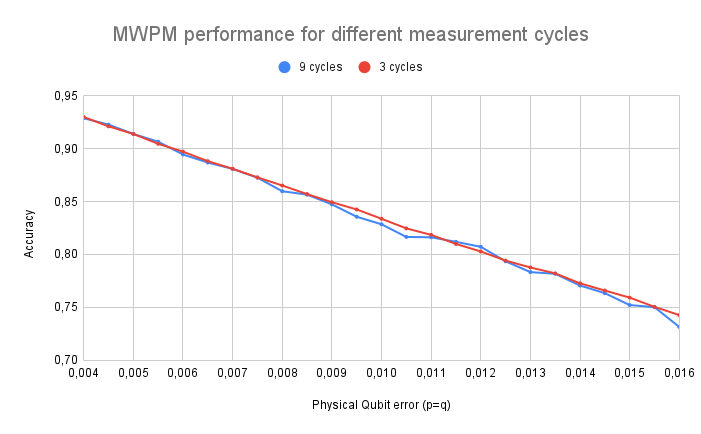}
\caption{Performance of MWPM for the depolarising plus measurement errors model using a different number or imperfect measurement cycles (3 and 9). The depolarising error probability ($p$) has been set equal to the measurement error probability ($q$). For this noise model a different number of measurement cycles does not modify significantly the decoding problem.}
\label{fig:cycles}
\end{figure}

In order to reproduce a more realistic error model it is necessary to include measurement errors and more correction cycles. In this section we will employ the following noise model: before a measurement cycle a depolarising noise is applied with single qubit error probability $p=0.01$. After applying the depolarising noise, an imperfect measurement cycle is performed. The measurement error probability has been set to $q=0.01$. Three rounds of depolarising error plus imperfect measurements are carried out on each surface code. This error model is different with respect to the channel noise model where gate errors are included. However, this noise model is easier to simulate, and can be employed for an initial study of the behaviour of high level decoders based on CNN, when measurement errors and many correction cycles are present. The MWPM algorithm still works for a circuit level noise model with many measurement cycles, in this case the error matches are carried out on a three dimensional graph~\cite{bib_noise1}. As benchmark, in this section, we employed the simplest version of the MWPM algorithm that considers all measurements as perfect. The performance of this algorithm is sub-optimal but can be improved by adjusting the weights in the matching graphs~\cite{autotune}.\\
It is important to notice that three measurement cycles is not the standard way to benchmark noise models with measurement errors, that is carried out using as many measurement cycles as the distance of the code. However, we have decided to reduce the number of measurement cycles in order to speed up the generation of the datasets thus obtaining more training samples. In fact, the time required to generate large training sets as well as their size was at the limit of the available computational resources, in particular in Sec.~\ref{sec_choice} where many training sets with different error probabilities have been employed.
Reducing the number of imperfect measurement cycles does not significantly change the decoding problem. As a reference, Fig.~\ref{fig:cycles} reports the performance of MWPM obtained on surface codes of dimension $d=9$ and the previously described noise model using three measurement cycles and nine measurement cycles.\\

\begin{table}
\caption{Best hyperparameters of the CNN for different code distances for a depolarising plus measurement errors model. $d$ is the code distance, CL is the number of convolutional layers, DL is the number of dense layers, N is the number of neurons in the dense layers and P is the total number of trainable weights.}
\label{tab:arch3}
\vspace{5mm}
\centering
\begin{tabular}{|c|c|c|c|c|}
\hline
$d$ & CL & DL & N & P \\
\hline
7 & 3 & 2 & 1024 & $6.4\times 10^6$ \\
9 & 3 & 2 & 1024 & $12.2\times 10^6$ \\
11 & 4 & 2 & 512 & $7.7\times 10^6$ \\
\hline
\end{tabular}
\end{table}

\begin{figure*}
\centering
\includegraphics[width=0.6\textwidth]{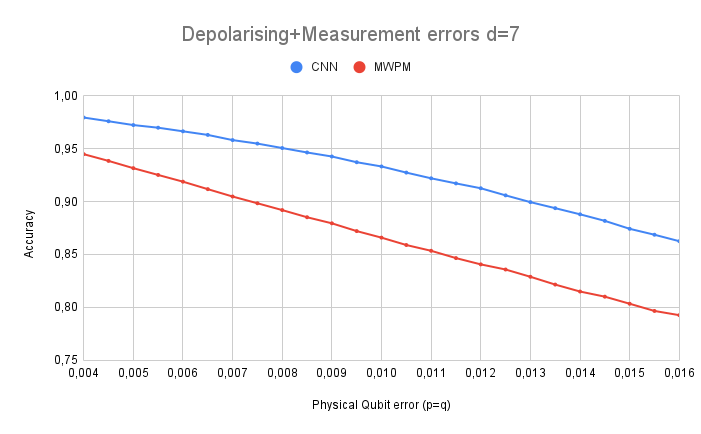}
\includegraphics[width=0.6\textwidth]{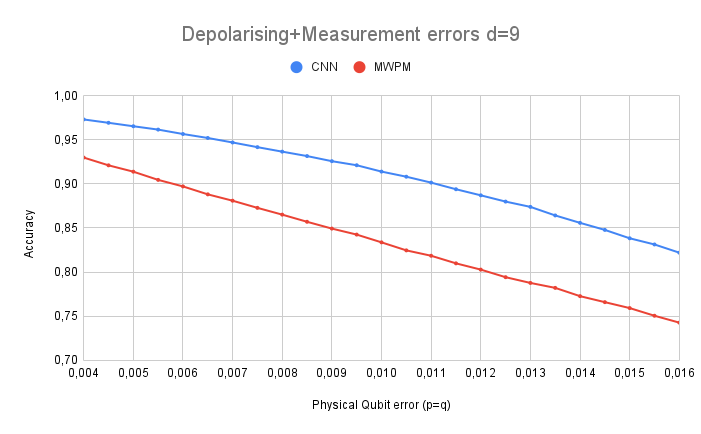}
\includegraphics[width=0.6\textwidth]{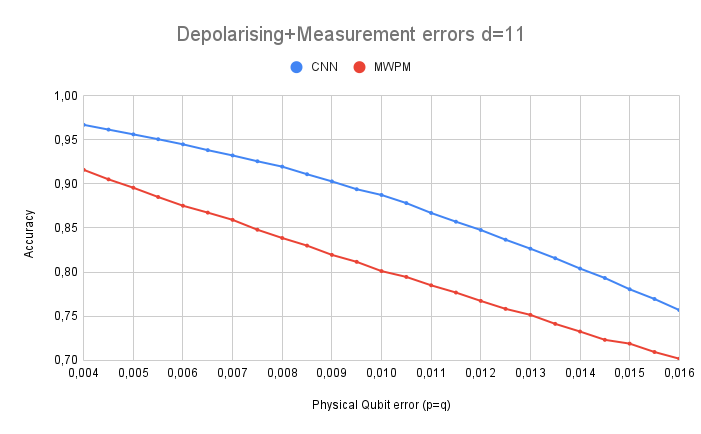}
\caption{Comparison between accuracy obtained with HLD based on CNN and the accuracy obtained with the classical decoder MWPM for the depolarising plus measurement errors model, the depolarising error probability ($p$) has been set equal to the measurement error probability ($q$).}
\label{tab:arch4}
\end{figure*}

For the depolarising plus measurement errors model it is necessary to find the best hyperparameters for the neural network employed in the HLD. All the characteristics of the neural networks, like kernel dimension and activation functions are the same employed for the depolarising noise. The hyperparameters that were changed to improve the performance regard the number of layers and neurons. The neural networks have been trained on datasets of $2\times 10^6$ elements, using $10\%$ of the original training set as the evaluation set.\\
For the CNN the best hyperparameters are reported in table~\ref{tab:arch3}. Some tests have been carried out using a FFNN, however no interesting results have been obtained using this architecture.
The accuracy obtained with the HLD, for the depolarising plus measurement errors model, is reported in Fig.~\ref{tab:arch4} and has been compared to the accuracy obtained with MWPM. The test on different single qubit error probabilities have been made on datasets of $2\times10^5$ elements. When measurement errors are included the accuracy of MWPM reduces sensibly while the HLD is able to better adapt to this different noise model.

\subsection{Dilated Convolution}\label{sec_dilated}

\begin{figure}
\centering
\includegraphics[width=0.5\textwidth]{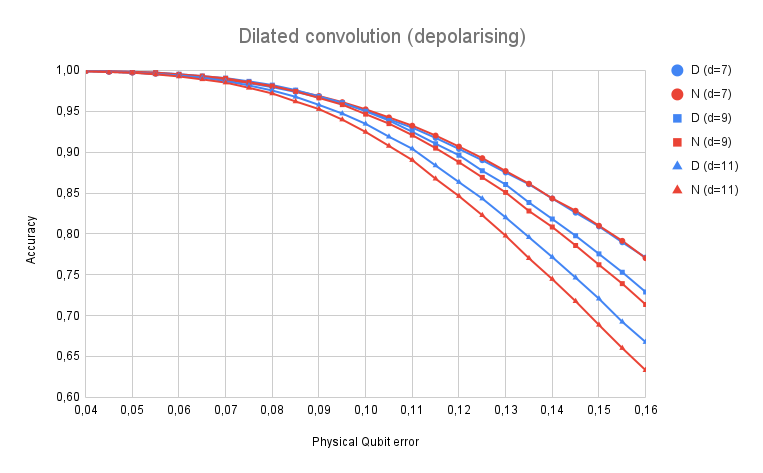}
\includegraphics[width=0.5\textwidth]{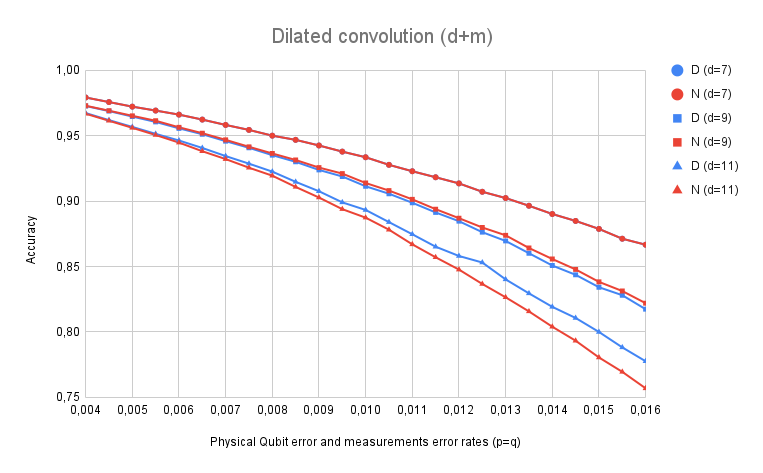}
\caption{Comparison of the accuracy, on different code distances ($d$), obtained with a CNN with no dilation factor (N) and the same model with a dilated convolution with a dilation rate equal to two on all the convolutional layers except the first one (D). For the depolarising error model (Top) the neural networks have been trained on a dataset with $p=0.1$. While for the depolarising plus measurement errors model (Bottom) $p=q=0.01$ (three measurement cycles). The graphs report the accuracy of the CNN decoder obtained using the described neural network architectures for different physical qubit error rates.}
\label{tab:dil}
\end{figure}

In order to improve the performance of the HLD for high distance codes, it should be useful to increase the local receptive field without incrementing the number of weights of the kernels. This result can be obtained with a dilated convolution~\cite{bib_dilated}. Three different implementations of a CNN with a dilated convolution have been tested. In the first case only the first layer of the CNN has a dilation factor equal to two. In the second case all the convolutional layers except the first one have dilation factor two and in the third case all the layers have this dilation factor. The training has been performed, for all code dimensions, on the same datasets described in Sec.~\ref{sec_dep} for the depolarising error model and in Sec.~\ref{sec_ddep} for the depolarising plus measurement errors model. The best results have been obtained using a dilated convolution for all the convolutional layers except the first one. The performance of this CNN architecture has been compared with the same model without the dilation factor, the results are reported in Fig.~\ref{tab:dil}.\\

The results obtained in this study show that dilated convolution may be a good way to improve the performance of the decoder for high distance codes. In fact, while for codes of distance 7 and 9 the use of dilated convolution doesn't alter the performance sensibly, for codes of distance 11, where there are more convolutional layers, it is possible to obtain a performance improvement. The performance increase obtained with the dilated convolution is due to a reduction of the number of trainable weights. In fact, no padding is used in the layers after the first one, so incrementing the local receptive field reduces the number of neurons of the first dense layer.

\section{Choice of the training set}\label{sec_choice}

\begin{figure*}
\centering
\includegraphics[width=0.6\textwidth]{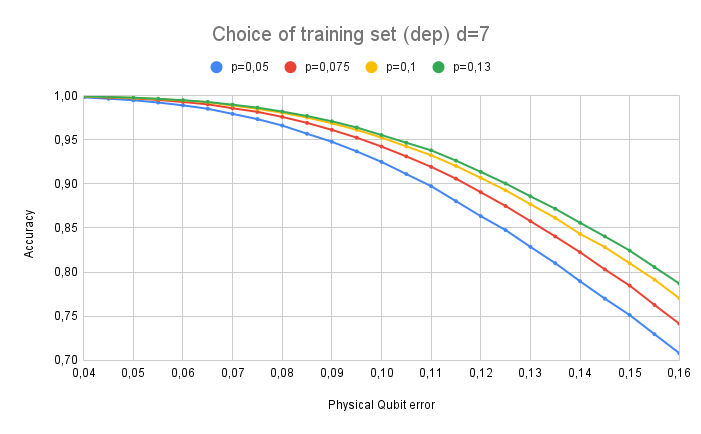}
\includegraphics[width=0.6\textwidth]{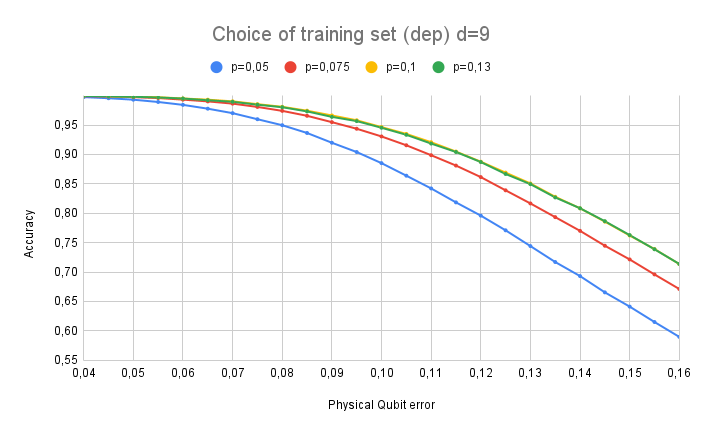}
\includegraphics[width=0.6\textwidth]{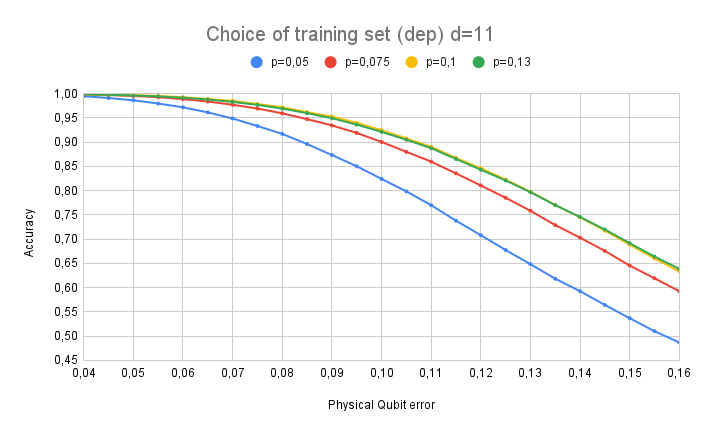}
\caption{Accuracy of the HLD trained and tested on different single qubit error probabilities for different code distances $d$. The training set single qubit error probability ($p$) is reported in the legend.}
\label{tab:stabdep}
\end{figure*}

\begin{figure*}
\centering
\includegraphics[width=0.6\textwidth]{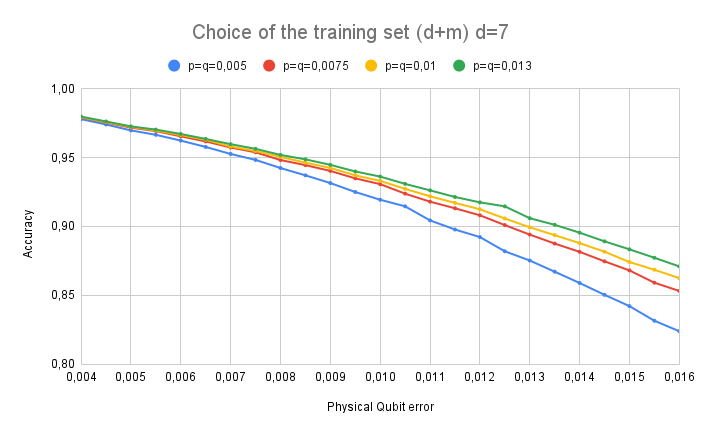}
\includegraphics[width=0.6\textwidth]{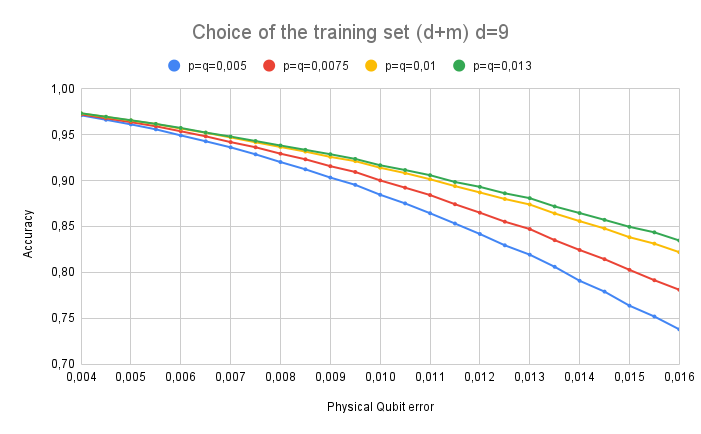}
\includegraphics[width=0.6\textwidth]{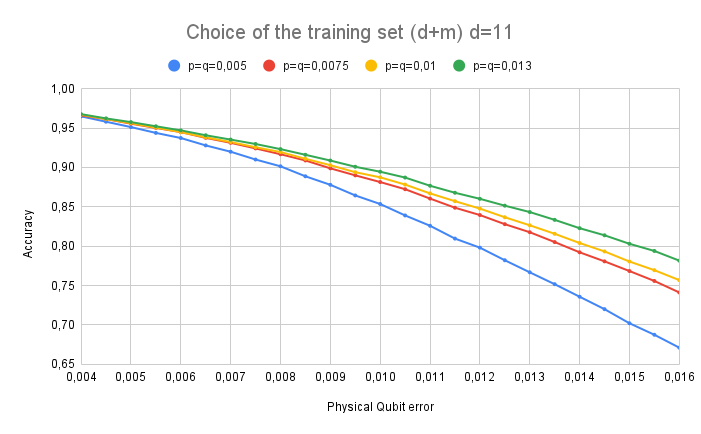}
\caption{Accuracy of the HLD trained and tested on different depolarising and measurement error probabilities for different code distances $d$. The training set single qubit error probability ($p$) and measurement error probability ($q$) is reported in the legend.}
\label{tab:stabdep2}
\end{figure*}

Neural networks are algorithms highly employed for their good generalisation properties. For the HLD it is possible to test these properties by training and testing the neural networks on datasets with different error probabilities. This is also very useful in real applications where the exact physical error rate is unknown.\\
For the depolarising error model, error probabilities of $p=0.05;0.075;0.1;0.13$ have been used to train the CNN for each code dimension. Each dataset is composed of $5\times10^6$ elements. 
The hyperparameters of the CNN are the ones described in Sec.~\ref{sec_ddep} (table~\ref{tab:arch}).
The performance has been evaluated on different error probabilities, the same datasets used in Sec.~\ref{sec_dep}.
The results of this study are reported in Fig.~\ref{tab:stabdep}.
Figure~\ref{tab:stabdep} shows that a better performance is obtained when the neural network is trained on a higher error probability. This is due to the fact that, in datasets with higher error probability there are statistically relevant samples that are not present in datasets with a lower error probability.
However, when the dimension of the code becomes greater, a high error probability for the training set may cause problems for the training of the neural network. This happened for codes of dimension $11$, using an error probability $p=0.13$, in this case it has been possible to train the network only by starting from the weights of the network previously trained on an error probability $p=0.1$. This procedure is very interesting as in this way it is possible to add more complicated error syndromes to the training set thus improving the performance of the decoder.\\
A similar analysis has been carried out for the depolarising plus measurement errors noise model. Four datasets ($2\times 10^6$ elements each) with different error probabilities have been generated for different code dimensions. The depolarising error probability is indicated with $p$ while the measurement error probability with $q$, the four datasets have error probabilities of: $p=q=0.005$, $p=q=0.0075$, $p=q=0.01$ and $p=q=0.013$.\\
For each code distance four neural networks have been trained using a single dataset. The CNN architecture and characteristics are the same of Sec.~\ref{sec_ddep} (table~\ref{tab:arch3}). The performance test with different error probabilities has been carried out on the same datasets used in Sec.~\ref{sec_ddep}. The results of this study are reported in Fig.~\ref{tab:stabdep2}, they are similar to the results obtained for the depolarising error model: it is better to use a higher error probability to train the neural network. However, also in this case, a high error probability may cause the neural network not to train correctly. In fact for both codes of distance 9 and 11 it was possible to train the network on an error probability $p=q=0.013$ only by starting from the weights of the neural network previously trained on a lower error probability.

\section{Model explainability}\label{sec_exp}

In order to trust complex and less transparent algorithms like artificial neural networks, it is necessary to know why they fail or work correctly. In particular, for convolutional neural networks trained for classification, it is important to understand the inputs that influence more the final decision. Many methods have been developed to construct saliency maps of the input pixels for image classification~\cite{bib_expML}. For example, with the GradCAM and Occlusion algorithms~\cite{bib_gradcam, bib_occlusion}, it is possible to obtain an heatmap of the input pixels based on their relevance for the output probability.\\
In the occlusion method several small regions of the image are systematically masked (all input pixels of the covered region are set to zero), and the changes in the loss function between the the occluded and standard image are recorded. 
The saliency value corresponding to a given masked region is given by:
\begin{multline*}
    Saliency(\text{region})=\\
    \left[loss(image_{\text{original}})-loss(image_{\text {masked}})\right]^2
\end{multline*}
We expect that a region that contributes more to the prediction of the model would change significantly the loss if masked. By shifting the masked region horizontally and vertically and repeating the process it is possible to construct the saliency map of the input features. For example, giving an image of 28x28 pixels and a occlusion patch of size 4x4 to mask the image, with a stride of 4 steps, an occlusion saliency map of size 7x7 is obtained. 
The salience map build in this way is then zoomed back to the original image resolution and overlaid to it.
Saliency maps help understanding if the trained HLD is performing as expected, and so allow to validate the algorithm, but they can also employed to better understand the errors of HLD in order to try to improve their performance. An original example of using salience maps to improve the algorithm itself is presented in Sec.~\ref{sec_improve}.

\subsection{Saliency map analysis}
\begin{figure}
\includegraphics[width=0.5\textwidth]{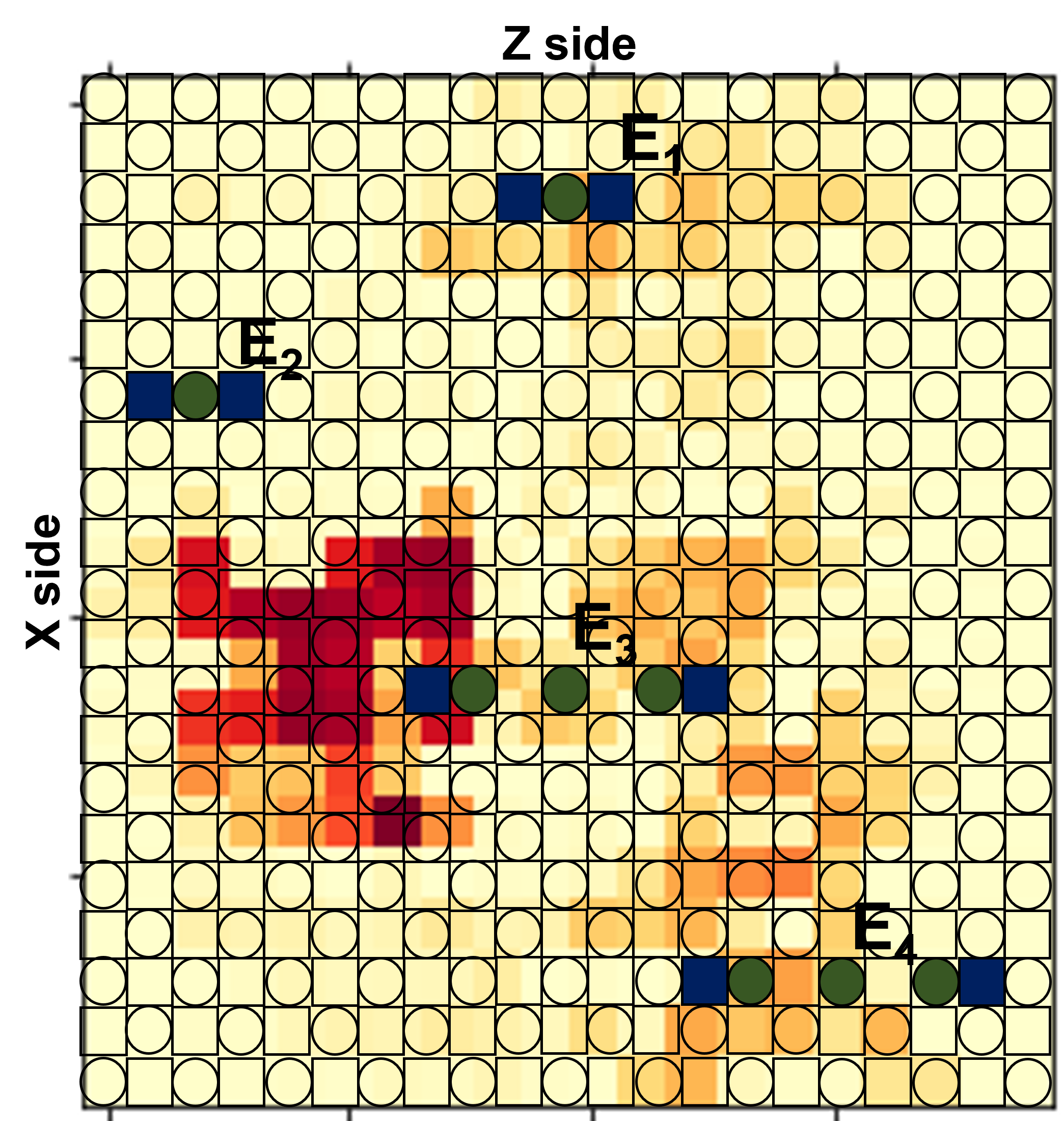}
\centering
\caption{Saliency map of a manually generated error syndrome analysed with a neural network decoder, the dark red regions are the ones that contribute more to the output. The error syndrome has been overlaid to the map, data qubits with $X$ errors are highlighted in green, while the measurement qubits that report the error signal are highlighted in blue.
Two single $X$ errors, $E_1$ and $E_2$, and two errors chains composed of three $X$ errors, $E_3$ and $E_4$ have been inserted in the surface code. The errors closer to the centre between the $X$ sides ($E_1$ and $E_3$) are the ones that contribute more for the classification.}
\label{fig:heat1}
\end{figure}

Occlusion method has been applied to the neural network of the HLD in order to understand how error syndromes are detected. Tests have been carried out for different code distances and neural networks trained on different error probabilities, the results obtained are very similar. In the following saliency maps of the input are reported for a depolarising noise model for codes of distance 11, using the neural network trained on an error probability $p=0.13$. In fact this is the neural network that showed better results for codes of dimension 11, and with the depolarising noise model it is easier to find the more critical regions of the input. In the implementation of the occlusion method squared occlusion patches of the input of dimension $2\times 2$ have been artificially set to zero. The dimension of the occluded window has been chosen in order to obtain a compromise between the smoothness and the granularity of the saliency map.\\
The first study regards how single errors or small chains of errors affect the final result, based on their position in the surface code (Fig.~\ref{fig:heat1}).
Two single $X$ errors have been placed, one in the middle between the $X$ sides ($E_1$), and one near an $X$ side ($E_2$). The single decoder is able to correct $E_2$ but creates an $X$ logical error when trying to decode $E_1$. The Neural network is able to identify the introduction of a logical error, the region near $E_1$ contributes to the output probability while the region near $E_2$ doesn't contribute significantly. This is due to the fact that occluding the region of the error $E_1$ reverses the output of the neural network ($X$ logical error can't be identified) thus significantly changing the loss.
Two errors chains composed of three $X$ errors have been placed one in the middle between the $X$ sides ($E_3$) and one near an $X$ side ($E_4$). The simple decoder is able to correct $E_4$ but creates a logical error correcting $E_3$. The regions near the ends of the chain $E_3$ contribute more to the output than the regions near the ends of the chain $E_4$. Also in this case this is due to the fact that covering the error syndrome of $E3$ reverse the output of the CNN. Moreover, we have observed also in other examples, that errors placed near the centre of the surface code influence more the output of the decoder.\\
The left side of the chain $E_3$ contributes more to the output than the right side. Similar asymmetries can be found in other saliency maps and change with respect to the horizontal or vertical position of the errors chain as well as the presence and position of other errors in the code. For example in the case reported in Fig.~\ref{fig:heat1} the asymmetry is mainly due to the presence of the error $E_2$. Note that there are active regions of the input between the right side of the chain $E_3$ and the left side of the chain $E_4$.
This first study considered only $X$ depolarising errors, similar results can be obtained for single $Z$ errors or small chains of $Z$ errors. In this case the errors that contribute more to the output of the decoder are the ones placed near the middle between the $Z$ sides.
\begin{figure}
\includegraphics[width=0.5\textwidth]{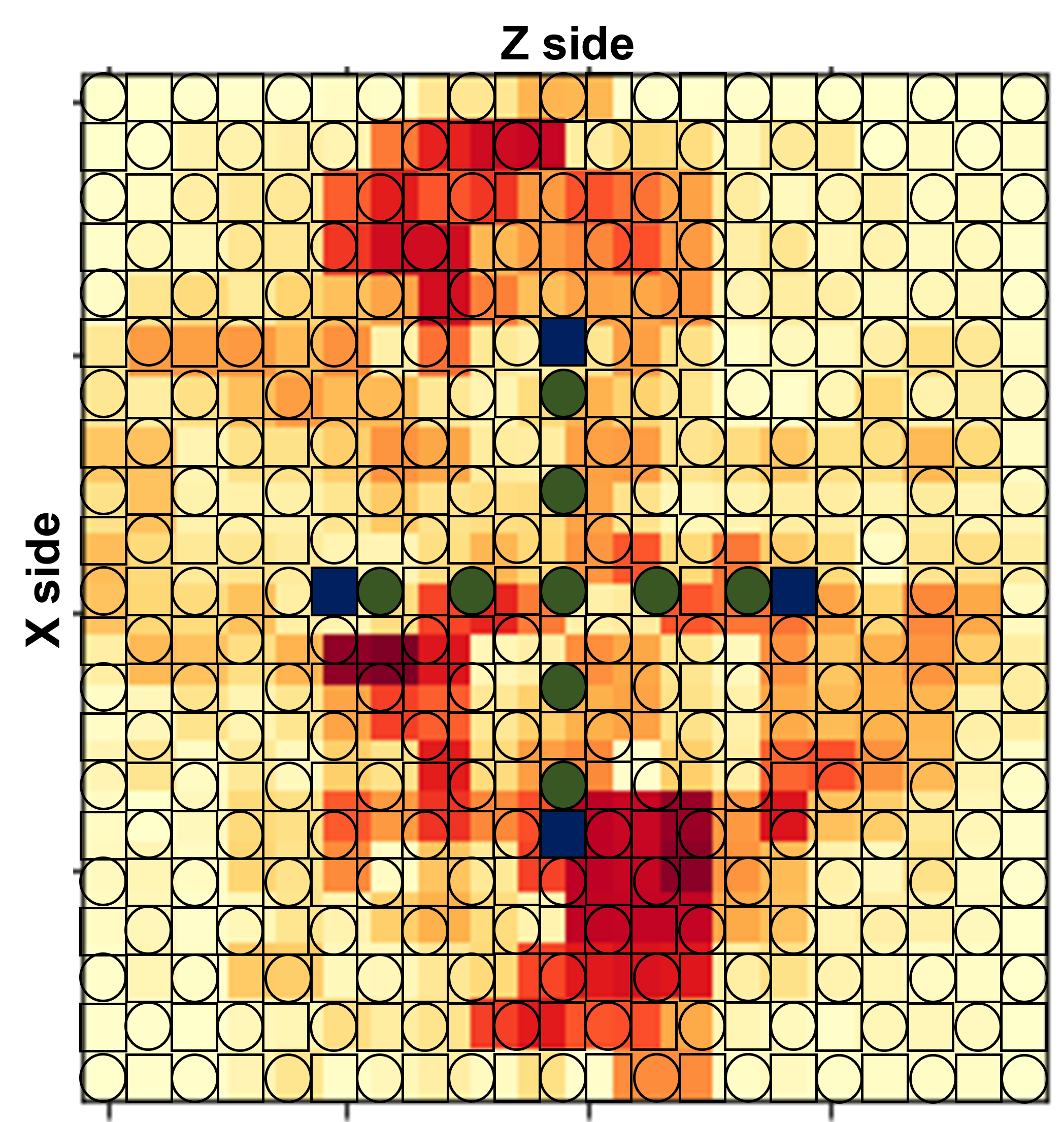}
\centering
\caption{Saliency map of a manually generated error syndrome analysed with a neural network decoder, the dark red regions are the ones that contribute more to the output. The error syndrome has been overlaid to the map, data qubit with depolarising errors are highlighted in green, while the measurement qubit that report the error signal are highlighted in blue.
Two errors chains composed respectively of five $X$ (horizontal) and $Z$ (vertical) errors have been inserted in the middle of the surface code. The HLD is not able to identify the creation of an $Y$ logical error after the correction of the simple decoder.}
\label{fig:heat2}
\end{figure}\\

For a depolarising error model the minimum weight perfect matching showed a better accuracy than the high level decoder, especially for codes of dimension 11. This means that there are cases where MWPM correctly identifies error chains while the high level decoder fails. An example is reported in Fig.~\ref{fig:heat2} where two error chains, composed of five errors, have been inserted in the code. The first chain is composed of $X$ errors and is placed horizontally in the centre of the code, the second chain, composed of $Z$ errors, is placed vertically in the centre of the code.
Chains of this kind are very dangerous as they may be easily misidentified by a decoder and corrected by matching the error syndrome with the nearest border of the code thus creating a logical error. This is the case of the HLD, that is not able to identify the creation of an $Y$ logical error after the correction of the simple decoder. In particular, the neural network classifies the error syndrome with no logical errors, this means that it is not able to find both the $X$ and $Z$ errors chains. Looking at the saliency map, reported in Fig.~\ref{fig:heat2}, it is possible to notice that the most relevant regions of the input are the ones located between the extremity of the chains and the nearest border of the surface code. This is probably due to the fact that the neural network decoder is looking for other errors in these regions in order to find the other extremities of the chains rather than matching $X$ and $Z$ syndromes with a single chain. This kind of error occurs almost every time a chain of length more than five appears in codes of distance 11. For codes of distance 9 and 7 the HLD correctly identifies chains of length 4 and 3 respectively. The MWPM is able to correct errors chains of length less than or equal $(d-1)/2$, where $d$ is the dimension of the code. This means that this algorithm can correct chains of length 5 for $d=11$ surface codes, while the maximum number of correctable errors in the chains is equal to 4 for $d=9$ and 3 for $d=7$. This explains the performance reduction of the HLD with respect to MWPM for codes of distance 11. The introduction of more samples in the training set, containing error chains of length five, can be a solution to improve the performance of HLD. This has been studied in Sec.~\ref{sec_improve}.\\

\begin{figure}
\includegraphics[width=0.5\textwidth]{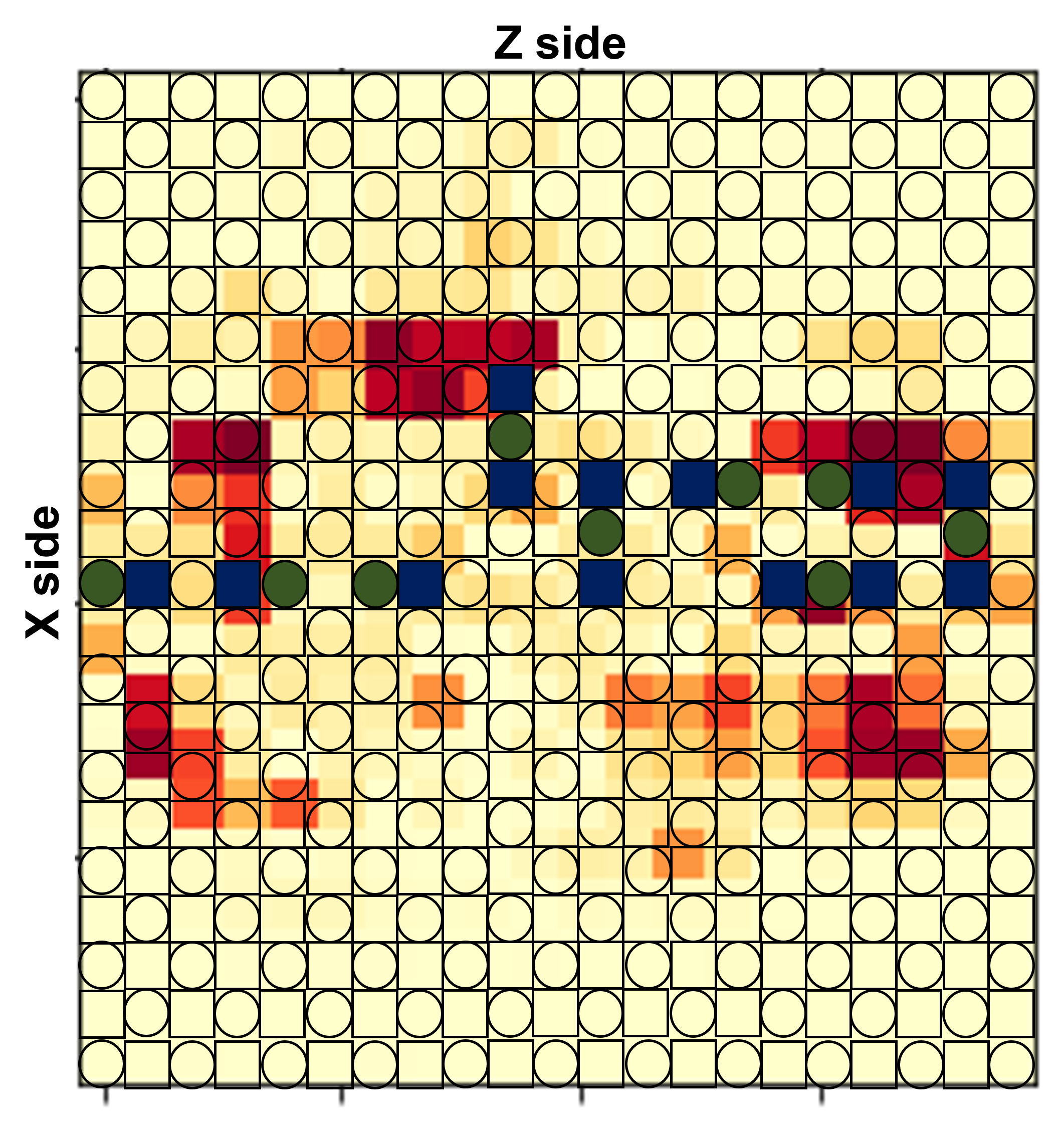}
\centering
\caption{Saliency map of an error syndrome analysed with a neural network decoder, the dark red regions are the ones that contribute more to the output. The error syndrome has been overlaid to the map, data qubit with $X$ errors are highlighted in green, while the measurement qubit that report the error signal are highlighted in blue. In this case the HLD is able to correct the error syndrome while the MWPM fails. This example has been generated using a depolarising error model (perfect measurements) with single qubit error probability $p=0.15$, the error chain responsible for the failure of MWPM has been extracted and analysed.}
\label{fig:heat3}
\end{figure}

There are some cases where the HLD is able to correct an error syndrome of a depolarising noise model where MWPM fails. An example is reported in Fig.~\ref{fig:heat3}, the error chain that gives problems to MWPM has been isolated by eliminating the other errors present in the code. After the correction of MWPM, an $X$ error chain connecting the $X$ sides of the code is introduced. On the other way, the HLD identifies all the logical errors introduced by the simple decoder. Looking at the saliency map in Fig.~\ref{fig:heat3} it is possible to notice that the central region of the surface code, where two isolated errors are present, is not really relevant for the output. On the other way, the region near the right $X$ side, where there is a small error chain, is the one that mostly affects the output.

\subsection{Performance enhancement with data augmentation driven by model explainability}\label{sec_improve}

\begin{figure*}
\includegraphics[width=0.7\textwidth]{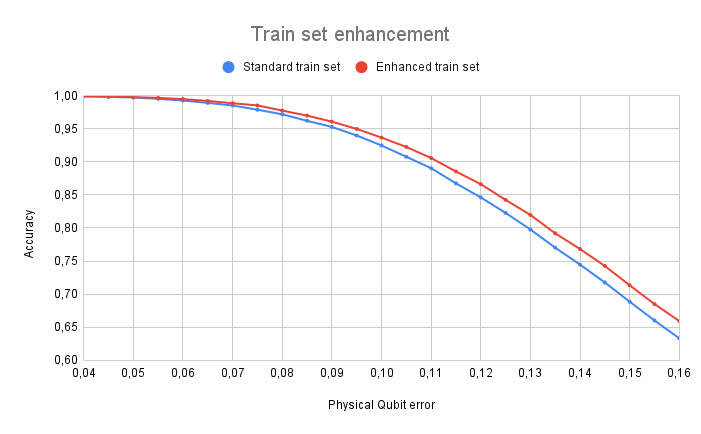}
\centering
\caption{Performance comparison between HLD trained on a standard train set with single qubit error probability $p=0.1$ and the model trained on the enhanced train set that includes samples of error chains of length five.}
\label{fig:enh}
\end{figure*}

Model Explainability may be an interesting tool to improve the performance of ML algorithms. This can be done, for example, by analysing the critical issues in order to remove or attenuate them. In this section we have applied this idea to improve the performance of HLD. Figure~\ref{fig:heat2} showed that the CNN decoder fails to recognise some error chains of length $5$ in a surface code of dimension $11$. As these kinds of error chains can be corrected by MWPM, they account for the better performance of this decoder with respect to a HLD. In order to reduce this problem, we have generated an enhanced augmented data training set that includes some samples of these error chains. These special samples have been generated with the following procedure.
An error chain, composed of five single qubit errors of the same kind and on the same row or column, is added to the surface code. Both the row (or column) position of the chain, as well as the error locations, are randomly drawn. A single $X$ error chain is added with probability $1/3$; with the same probability a single $Z$ error chain is added and, in the other cases, both an $X$ and a $Z$ error chains are added.
The enhanced train set used to improve the performance of the CNN is composed of $7\times 10^6$ samples divided as follows. 
$10^6$ special samples containing error chains of length $5$ (as previously described).
$5\times10^6$ samples with a single qubit error probability $p=0.1$ (standard train set employed in Sec.~\ref{sec_dep}).
$10^6$ samples with a single qubit error probability $p=0.13$; a higher single qubit error probability helps produce samples with longer error chains.
The employed CNN is the same described in Sec.~\ref{sec_dep} for surface codes of dimension $11$. As the enhanced train set contains samples of different kind, the training batch size has been set to $128$ to increase training stability. The model has been trained for four epochs, starting from the parameters trained in Sec.~\ref{sec_dep} on the standard train set with a single qubit error probability $p=0.1$. The trained model has been tested on different single qubit error probabilities, the same test sets introduced in Sec.~\ref{sec_dep}.\\

\begin{figure}[t]
\includegraphics[width=0.5\textwidth]{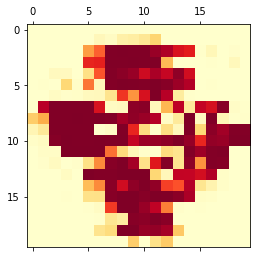}
\centering
\caption{Saliency map of the same error syndrome reported in Fig.~\ref{fig:heat2}, decoded with the HLD trained on the enhanced train set. The dark red regions are the ones that contribute more to the output, they are placed near the error signals and the centre of the surface code. This HLD is able to correctly identify the presence of error chains from a sparse error syndrome.}
\label{fig:enh_salience}
\end{figure}

Figure~\ref{fig:enh} reports the performance improvement obtained with the enhanced train set, with respect to the standard train set employed in Sec.~\ref{sec_dep}. The performance improvement is significant, about $1\%$ for single qubit error probability $p=0.1$ and up to $2\%$ for higher single qubit error probabilities. Moreover, the new model is able to correctly decode the example reported in Fig.~\ref{fig:heat2}. The new saliency map, obtained with this model, is reported in Fig.~\ref{fig:enh_salience}. A comparison with Fig.~\ref{fig:heat2} shows how the HLD trained on the enhanced dataset gives the same importance to $X$ and $Z$ error signals.
This decoder is able to better understand the importance of the error signals at the end of the error chains, as well as the central region of the surface code.

\section{Conclusions}
Neural network based decoders have proved to be excellent algorithms for the decoding of surface codes due to their constant execution time, good accuracy and adaptability to different noise models. The use of a convolutional architecture, with respect to a dense architecture, helps scaling to higher distance codes. However, as incrementing the distance of the code increments exponentially the possible error syndromes, larger datasets are required for correct training. This makes difficult to apply a simple convolutional architecture to decode really high distance codes. Convolutional neural networks decoders remain interesting as they may be used as the first step of a more sophisticated decoder to process local information~\cite{scalable}. Moreover, it is likely that in the future only small distance surface codes will be available for the first tests.\\

The results obtained in this work are promising and helpful to improve the accuracy obtained with HLD based on convolutional neural networks. Different convolutional neural network architectures and training strategies have been tested for different code distances and noise models. The results suggest that using a training set with a higher error probability helps improving the performance of the decoder, and that successful training benefits can be obtained from a pre-training based on examples on a lower error probability. It was also shown how it is possible to use dilated convolution to reduce the number of parameters that can be trained with decoders based on very deep convolutional neural networks. This suggest a possible way to try to scale the convolutional neural networks based decoder to larger surface codes.\\
Finally, explainability methods have been proposed and applied to get insights on how error syndromes are classified by the neural network decoder. A better knowledge of the causes of the failure of the decoding is in fact fundamental to improve the performance, robustness and confidence in neural network HLD for real applications. In this respect an original example of use of the salience maps to improve the HLD algorithm has been presented.\\

\noindent
\textbf{Acknowledgments}\\
This work has been partially supported by ICSC - Centro Nazionale di Ricerca in High Performance Computing, Big Data and Quantum Computing, founded by European Union - NextGeneration EU.
\newpage

\printbibliography

@article{bib_decoherence,
    title = {Quantum decoherence},
    journal = {Physics Reports},
    volume = {831},
    pages = {1-57},
    year = {2019},
    issn = {0370-1573},
    doi = {https://doi.org/10.1016/j.physrep.2019.10.001},
    url = {https://www.sciencedirect.com/science/article/pii/S0370157319303084},
    author = {Maximilian Schlosshauer},
}

@book{bib_principles2,
  title={Principles Of Quantum Computation And Information - Volume II: Basic Tools And Special Topics},
  author={Benenti, G. and Casati, G. and Strini, G.},
  isbn={9789814365550},
  url={https://books.google.it/books?id=Its7DQAAQBAJ},
  year={2007},
  address={Italy},
  publisher={World Scientific Publishing Company}
}

@article{bib_qecc1,
   title={Quantum error correction: an introductory guide},
   volume={60},
   ISSN={1366-5812},
   url={http://dx.doi.org/10.1080/00107514.2019.1667078},
   DOI={10.1080/00107514.2019.1667078},
   number={3},
   journal={Contemporary Physics},
   publisher={Informa UK Limited},
   author={Roffe, Joschka},
   year={2019},
   month={7},
   pages={226–245}
}

@article{bib_qecc2,
   title={Theory of Quantum Error Correction for General Noise},
   volume={84},
   ISSN={1079-7114},
   url={http://dx.doi.org/10.1103/PhysRevLett.84.2525},
   DOI={10.1103/physrevlett.84.2525},
   number={11},
   journal={Physical Review Letters},
   publisher={American Physical Society (APS)},
   author={Knill, Emanuel and Laflamme, Raymond and Viola, Lorenza},
   year={2000},
   month={3},
   pages={2525–2528}
}

@article{XZopt,
	doi = {10.1103/physreva.90.032326},
	url = {https://doi.org/10.11032Fphysreva.90.032326},
	year = 2014,
	month = {9},
	publisher = {American Physical Society ({APS})},
	volume = {90},
	number = {3},
	author = {Sergey Bravyi and Martin Suchara and Alexander Vargo},
	title = {Efficient algorithms for maximum likelihood decoding in the surface code},
	journal = {Physical Review A}
}

@article{bib_kitaev,
   title={Fault-tolerant quantum computation by anyons},
   volume={303},
   ISSN={0003-4916},
   url={http://dx.doi.org/10.1016/S0003-4916(02)00018-0},
   DOI={10.1016/s0003-4916(02)00018-0},
   number={1},
   journal={Annals of Physics},
   publisher={Elsevier BV},
   author={Kitaev, A.Yu.},
   year={2003},
   month={1},
   pages={2–30}
}

@article{bib_surface1,
   title={Surface codes: Towards practical large-scale quantum computation},
   volume={86},
   ISSN={1094-1622},
   url={http://dx.doi.org/10.1103/PhysRevA.86.032324},
   DOI={10.1103/physreva.86.032324},
   number={3},
   journal={Physical Review A},
   publisher={American Physical Society (APS)},
   author={Fowler, Austin G. and Mariantoni, Matteo and Martinis, John M. and Cleland, Andrew N.},
   year={2012},
   month={9}
}

@article{bib_decodespeed,
    title={Decoding surface code with a distributed neural network based decoder}, 
    author={Savvas Varsamopoulos and Koen Bertels and Carmen G. Almudever},
    year={2019},
    eprint={1901.10847},
    archivePrefix={arXiv},
    primaryClass={quant-ph}
}

@article{bib_vars0,
   title={Comparing Neural Network Based Decoders for the Surface Code},
   volume={69},
   ISSN={2326-3814},
   url={http://dx.doi.org/10.1109/TC.2019.2948612},
   DOI={10.1109/tc.2019.2948612},
   number={2},
   journal={IEEE Transactions on Computers},
   publisher={Institute of Electrical and Electronics Engineers (IEEE)},
   author={Varsamopoulos, Savvas and Bertels, Koen and Almudever, Carmen Garcia},
   year={2020},
   month={2},
   pages={300–311}
}

@book{bib_qiqc,
    author = {Nielsen, Michael A. and Chuang, Isaac L.},
    title = {Quantum Computation and Quantum Information: 10th Anniversary Edition},
    year = {2011},
    isbn = {1107002176},
    publisher = {Cambridge University Press},
    address = {USA},
    edition = {10th},
}

@article{bib_latticesurgery,
   title={Surface code quantum computing by lattice surgery},
   volume={14},
   ISSN={1367-2630},
   url={http://dx.doi.org/10.1088/1367-2630/14/12/123011},
   DOI={10.1088/1367-2630/14/12/123011},
   number={12},
   journal={New Journal of Physics},
   publisher={IOP Publishing},
   author={Horsman, Clare and Fowler, Austin G and Devitt, Simon and Meter, Rodney Van},
   year={2012},
   month={12},
   pages={123011}
}

@article{bib_lattice,
   title={A Game of Surface Codes: Large-Scale Quantum Computing with Lattice Surgery},
   volume={3},
   ISSN={2521-327X},
   url={http://dx.doi.org/10.22331/q-2019-03-05-128},
   DOI={10.22331/q-2019-03-05-128},
   journal={Quantum},
   publisher={Verein zur Forderung des Open Access Publizierens in den Quantenwissenschaften},
   author={Litinski, Daniel},
   year={2019},
   month={3},
   pages={128}
}

@article{bib_logicalop,
   title={Quantum error correction for quantum memories},
   volume={87},
   ISSN={1539-0756},
   url={http://dx.doi.org/10.1103/RevModPhys.87.307},
   DOI={10.1103/revmodphys.87.307},
   number={2},
   journal={Reviews of Modern Physics},
   publisher={American Physical Society (APS)},
   author={Terhal, Barbara M.},
   year={2015},
   month={4},
   pages={307–346}
}

@article{bib_noise1,
   title={Analytic asymptotic performance of topological codes},
   volume={87},
   ISSN={1094-1622},
   url={http://dx.doi.org/10.1103/PhysRevA.87.040301},
   DOI={10.1103/physreva.87.040301},
   number={4},
   journal={Physical Review A},
   publisher={American Physical Society (APS)},
   author={Fowler, Austin G.},
   year={2013},
   month={4}
}

@article{autotune,
	doi = {10.1103/physrevx.2.041003},
	url = {https://doi.org/10.11032Fphysrevx.2.041003},
	year = 2012,
	month = {10},
	publisher = {American Physical Society ({APS})},
	volume = {2},
	number = {4},
	author = {Austin G. Fowler and Adam C. Whiteside and Angus L. McInnes and Alimohammad Rabbani},
	title = {Topological Code Autotune},
	journal = {Physical Review X}
}

@article{bib_noise2,
    title={Threshold error rates for the toric and surface codes}, 
    author={D. S. Wang and A. G. Fowler and A. M. Stephens and L. C. L. Hollenberg},
    year={2009},
    eprint={0905.0531},
    archivePrefix={arXiv},
    primaryClass={quant-ph}
}

@article{bib_noise3,
   title={Low-distance surface codes under realistic quantum noise},
   volume={90},
   ISSN={1094-1622},
   url={http://dx.doi.org/10.1103/PhysRevA.90.062320},
   DOI={10.1103/physreva.90.062320},
   number={6},
   journal={Physical Review A},
   publisher={American Physical Society (APS)},
   author={Tomita, Yu and Svore, Krysta M.},
   year={2014},
   month={12}
}

@article{bib_hilbert,
    title={Stabilizer Codes and Quantum Error Correction}, 
    author={Daniel Gottesman},
    year={1997},
    eprint={9705052},
    archivePrefix={arXiv},
    primaryClass={quant-ph}
}

@article{bib_mwpm1,
   title={Adaptive Weight Estimator for Quantum Error Correction in a Time‐Dependent Environment},
   volume={1},
   ISSN={2511-9044},
   url={http://dx.doi.org/10.1002/qute.201800012},
   DOI={10.1002/qute.201800012},
   number={1},
   journal={Advanced Quantum Technologies},
   publisher={Wiley},
   author={Spitz, Stephen T. and Tarasinski, Brian and Beenakker, Carlo W. J. and O’Brien, Thomas E.},
   year={2018},
   month={7},
}

@article{bib_mwpm0,
    title={Minimum weight perfect matching of fault-tolerant topological quantum error correction in average $O(1)$ parallel time}, 
    author={Austin G. Fowler},
    year={2014},
    eprint={1307.1740},
    archivePrefix={arXiv},
    primaryClass={quant-ph}
}

@article{bib_mwpm2,
    author = {Cook, William and Rohe, André},
    title = {Computing Minimum-Weight Perfect Matchings},
    journal = {INFORMS Journal on Computing},
    volume = {11},
    number = {2},
    pages = {138-148},
    year = {1999},
    doi = {10.1287/ijoc.11.2.138},
    URL = {https://doi.org/10.1287/ijoc.11.2.138},
}

@article{rn2,
   title={Fast Decoders for Topological Quantum Codes},
   volume={104},
   ISSN={1079-7114},
   url={http://dx.doi.org/10.1103/PhysRevLett.104.050504},
   DOI={10.1103/physrevlett.104.050504},
   number={5},
   journal={Physical Review Letters},
   publisher={American Physical Society (APS)},
   author={Duclos-Cianci, Guillaume and Poulin, David},
   year={2010},
   month={2}
}

@article{optimal,
    title={Optimal complexity correction of correlated errors in the surface code},
    author={Austin G. Fowler},
    year={2013},
    eprint={1310.0863},
    archivePrefix={arXiv},
    primaryClass={quant-ph}
}

@article{corrxy,
   title={A decoding algorithm for CSS codes using the X/Z correlations},
   url={http://dx.doi.org/10.1109/ISIT.2014.6874997},
   DOI={10.1109/isit.2014.6874997},
   journal={2014 IEEE International Symposium on Information Theory},
   publisher={IEEE},
   author={Delfosse, Nicolas and Tillich, Jean-Pierre},
   year={2014},
   month={6}
}

@article{boltzmann,
   title={Neural Decoder for Topological Codes},
   volume={119},
   ISSN={1079-7114},
   url={http://dx.doi.org/10.1103/PhysRevLett.119.030501},
   DOI={10.1103/physrevlett.119.030501},
   number={3},
   journal={Physical Review Letters},
   publisher={American Physical Society (APS)},
   author={Torlai, Giacomo and Melko, Roger G.},
   year={2017},
   month={7}
}

@article{toric,
    author = {Krastanov, Stefan and Jiang, Liang},
    year = {2017},
    month = {05},
    pages = {},
    title = {Deep Neural Network Probabilistic Decoder for Stabilizer Codes},
    volume = {7},
    journal = {Scientific Reports},
    doi = {10.1038/s41598-017-11266-1}
}

@article{ensemble,
   title={Neural ensemble decoding for topological quantum error-correcting codes},
   volume={101},
   ISSN={2469-9934},
   url={http://dx.doi.org/10.1103/PhysRevA.101.032338},
   DOI={10.1103/physreva.101.032338},
   number={3},
   journal={Physical Review A},
   publisher={American Physical Society (APS)},
   author={Sheth, Milap and Jafarzadeh, Sara Zafar and Gheorghiu, Vlad},
   year={2020},
   month={3}
}

@article{scalable,
      title={Scalable Neural Decoder for Topological Surface Codes}, 
      author={Kai Meinerz and Chae-Yeun Park and Simon Trebst},
      year={2021},
      eprint={2101.07285},
      archivePrefix={arXiv},
      primaryClass={quant-ph}
}

@article{large1,
   title={Neural Network Decoders for Large-Distance 2D Toric Codes},
   volume={4},
   ISSN={2521-327X},
   url={http://dx.doi.org/10.22331/q-2020-08-24-310},
   DOI={10.22331/q-2020-08-24-310},
   journal={Quantum},
   publisher={Verein zur Forderung des Open Access Publizierens in den Quantenwissenschaften},
   author={Ni, Xiaotong},
   year={2020},
   month={8},
   pages={310}
}

@article{large2,
   title={General framework for constructing fast and near-optimal machine-learning-based decoder of the topological stabilizer codes},
   volume={2},
   ISSN={2643-1564},
   url={http://dx.doi.org/10.1103/PhysRevResearch.2.033399},
   DOI={10.1103/physrevresearch.2.033399},
   number={3},
   journal={Physical Review Research},
   publisher={American Physical Society (APS)},
   author={Davaasuren, Amarsanaa and Suzuki, Yasunari and Fujii, Keisuke and Koashi, Masato},
   year={2020},
   month={9}
}

@article{assisted,
   title={Machine-learning-assisted correction of correlated qubit errors in a topological code},
   volume={2},
   ISSN={2521-327X},
   url={http://dx.doi.org/10.22331/q-2018-01-29-48},
   DOI={10.22331/q-2018-01-29-48},
   journal={Quantum},
   publisher={Verein zur Forderung des Open Access Publizierens in den Quantenwissenschaften},
   author={Baireuther, Paul and O’Brien, Thomas E. and Tarasinski, Brian and Beenakker, Carlo W. J.},
   year={2018},
   month={1},
   pages={48}
}

@article{bib_vars1,
   title={Decoding small surface codes with feedforward neural networks},
   volume={3},
   ISSN={2058-9565},
   url={http://dx.doi.org/10.1088/2058-9565/aa955a},
   DOI={10.1088/2058-9565/aa955a},
   number={1},
   journal={Quantum Science and Technology},
   publisher={IOP Publishing},
   author={Varsamopoulos, Savvas and Criger, Ben and Bertels, Koen},
   year={2017},
   month={11},
}

@article{bib_dilated,
      title={Multi-Scale Context Aggregation by Dilated Convolutions}, 
      author={Fisher Yu and Vladlen Koltun},
      year={2016},
      eprint={1511.07122},
      archivePrefix={arXiv},
      primaryClass={cs.CV}
}

@article{bib_expML,
      title={Object Detectors Emerge in Deep Scene CNNs}, 
      author={Bolei Zhou and Aditya Khosla and Agata Lapedriza and Aude Oliva and Antonio Torralba},
      year={2015},
      eprint={1412.6856},
      archivePrefix={arXiv},
      primaryClass={cs.CV}
}

@article{bib_gradcam,
   title={Grad-CAM: Visual Explanations from Deep Networks via Gradient-Based Localization},
   volume={128},
   ISSN={1573-1405},
   url={http://dx.doi.org/10.1007/s11263-019-01228-7},
   DOI={10.1007/s11263-019-01228-7},
   number={2},
   journal={International Journal of Computer Vision},
   publisher={Springer Science and Business Media LLC},
   author={Selvaraju, Ramprasaath R. and Cogswell, Michael and Das, Abhishek and Vedantam, Ramakrishna and Parikh, Devi and Batra, Dhruv},
   year={2019},
   month={10},
   pages={336–359}
}

@article{bib_occlusion,
      title={Visualizing and Understanding Convolutional Networks}, 
      author={Matthew D Zeiler and Rob Fergus},
      year={2013},
      eprint={1311.2901},
      archivePrefix={arXiv},
      primaryClass={cs.CV}
}

@article{bib_mld,
   title={Efficient algorithms for maximum likelihood decoding in the surface code},
   volume={90},
   ISSN={1094-1622},
   url={http://dx.doi.org/10.1103/PhysRevA.90.032326},
   DOI={10.1103/physreva.90.032326},
   number={3},
   journal={Physical Review A},
   publisher={American Physical Society (APS)},
   author={Bravyi, Sergey and Suchara, Martin and Vargo, Alexander},
   year={2014},
   month={9}
}

@article{bib_rg,
      title={A renormalization group decoding algorithm for topological quantum codes}, 
      author={Guillaume Duclos-Cianci and David Poulin},
      year={2010},
      eprint={1006.1362},
      archivePrefix={arXiv},
      primaryClass={quant-ph}
}

@article{bib_ca,
   title={Cellular-automaton decoders for topological quantum memories},
   volume={1},
   ISSN={2056-6387},
   url={http://dx.doi.org/10.1038/npjqi.2015.10},
   DOI={10.1038/npjqi.2015.10},
   number={1},
   journal={npj Quantum Information},
   publisher={Springer Science and Business Media LLC},
   author={Herold, Michael and Campbell, Earl T and Eisert, Jens and Kastoryano, Michael J},
   year={2015},
   month={10}
}

@article{bib_mcmc,
   title={Efficient Markov chain Monte Carlo algorithm for the surface code},
   volume={89},
   ISSN={1094-1622},
   url={http://dx.doi.org/10.1103/PhysRevA.89.022326},
   DOI={10.1103/physreva.89.022326},
   number={2},
   journal={Physical Review A},
   publisher={American Physical Society (APS)},
   author={Hutter, Adrian and Wootton, James R. and Loss, Daniel},
   year={2014},
   month={2}
}

@article{bib_new2021,
    title={Efficient Decoding of Surface Code Syndromes for Error Correction in Quantum Computing}, 
    author={Debasmita Bhoumik and Pinaki Sen and Ritajit Majumdar and Susmita Sur-Kolay and Latesh Kumar K J and Sundaraja Sitharama Iyengar},
    year={2021},
    eprint={2110.10896},
    archivePrefix={arXiv},
    primaryClass={quant-ph}
}

@article{ch1,
   title={Techniques for combining fast local decoders with global decoders under circuit-level noise},
   volume={8},
   ISSN={2058-9565},
   url={http://dx.doi.org/10.1088/2058-9565/ace64d},
   DOI={10.1088/2058-9565/ace64d},
   number={4},
   journal={Quantum Science and Technology},
   publisher={IOP Publishing},
   author={Chamberland, Christopher and Goncalves, Luis and Sivarajah, Prasahnt and Peterson, Eric and Grimberg, Sebastian},
   year={2023},
   month=jul, pages={045011} }

@article{ch2,
   title={Deep neural decoders for near term fault-tolerant experiments},
   volume={3},
   ISSN={2058-9565},
   url={http://dx.doi.org/10.1088/2058-9565/aad1f7},
   DOI={10.1088/2058-9565/aad1f7},
   number={4},
   journal={Quantum Science and Technology},
   publisher={IOP Publishing},
   author={Chamberland, Christopher and Ronagh, Pooya},
   year={2018},
   month=jul, pages={044002} }

\end{document}